\documentclass[twocolumn]{aastex631}

\usepackage[english]{babel}
\usepackage{amsmath}
\usepackage{amssymb}
\usepackage{graphicx}
\usepackage{subfigure}
\usepackage[normalem]{ulem}
\usepackage{enumerate}
\usepackage{booktabs}

\usepackage{verbatim}

\usepackage{color}
\usepackage{multirow}
\usepackage{mathtools}
\usepackage{epstopdf}
\usepackage{tabularx}

\usepackage{url}
\usepackage{xcolor}

\def\apjl{ApJL}
\def\apjs{ApJS}
\def\aap{A\&A}

\makeatletter
\def\@hex@@Hex#1%
 {\if a#1A\else \if b#1B\else \if c#1C\else \if d#1D\else
  \if e#1E\else \if f#1F\else #1\fi\fi\fi\fi\fi\fi \@hex@Hex}
\makeatother

\definecolor{apcolor}{HTML}{b3003b}
\definecolor{cbcolor}{HTML}{ff0f00}
\definecolor{afcolor}{HTML}{b3443c}
\definecolor{vgcolor}{HTML}{8F00FF}
\definecolor{tbdcolor}{HTML}{E8A95E}
\definecolor{stefcolor}{HTML}{0047ab}

\graphicspath{{plots_png/},{plots/}}

\shorttitle{MW progenitors with JWST}
\shortauthors{Rusta, Salvadori et al.}

\begin{document}

\title{Linking high-$z$ and low-$z$: Are We Observing the Progenitors of the Milky Way with JWST?}

\correspondingauthor{Elka Rusta}
\email{elka.rusta@unifi.it}

\author[0009-0006-4326-6097]{Elka Rusta}
\author[0000-0001-7298-2478]{Stefania Salvadori}
\affiliation{Dipartimento di Fisica e Astronomia, Università degli Studi di Firenze, Largo E. Fermi 1, 50125, Firenze, Italy}
\affiliation{INAF/Osservatorio Astrofisico di Arcetri, Largo E. Fermi 5, 50125, Firenze, Italy}

\author[0000-0001-5487-0392]{Viola Gelli}
\affiliation{Cosmic Dawn Center (DAWN), Denmark}
\affiliation{Niels Bohr Institute, University of Copenhagen, Jagtvej 128, 2200 Copenhagen N, Denmark}

\author[0000-0002-3524-7172]{Ioanna Koutsouridou}
\author[0000-0002-9889-4238]{Alessandro Marconi}
\affiliation{Dipartimento di Fisica e Astronomia, Università degli Studi di Firenze, Largo E. Fermi 1, 50125, Firenze, Italy}
\affiliation{INAF/Osservatorio Astrofisico di Arcetri, Largo E. Fermi 5, 50125, Firenze, Italy}

\begin{abstract}

The recent JWST observation of the \textit{Firefly Sparkle} at $z=8.3$ offers a unique opportunity to link the high- and the low-$z$ Universe. Indeed, the claim of it being a Milky Way (MW) type of assembly at the cosmic dawn opens the possibility of interpreting the observation with locally calibrated galaxy-formation models. Here, we use the MW-evolution model NEFERTITI to perform forward modeling of our Galaxy’s progenitors at high-$z$. We build a set of mock spectra for the MW building blocks to make predictions for JWST and to interpret the Firefly Sparkle observation. 
First, we find that the most massive MW progenitor becomes detectable in a deep survey like JADES from $z\approx 8.2$, meaning that we could have already observed MW-analogs that still need interpretation. Second, we provide predictions for the number of detectable MW progenitors in lensed surveys like CANUCS, and interpret the Firefly Sparkle as a group of MW building blocks. Both the number of detections and the observed NIRCam photometry are consistent with our predictions. By identifying the MW progenitors whose mock photometry best fits the data, we find bursty and extended star-formation histories, lasting $> 150-300$~Myr, and estimate their properties: $M_h \approx 10^{8-9} \, M_{\odot}$, $ M_\star \approx 10^{6.2-7.5}\, M_{\odot}$, $ SFR \approx 0.04-0.20 \, M_{\odot} yr^{-1}$ and $ Z_{gas} \approx 0.04 - 0.24 \, Z_{\odot}$. 
Uncovering the properties of MW-analogs at cosmic dawn by combining JWST observations and locally-constrained models, will allow us to understand our Galaxy’s formation, linking the high- and low-$z$ perspectives.

\end{abstract}

\keywords{galaxies: high-redshift --- galaxies: evolution --- galaxies: formation}

\section{Introduction}
\label{sec:intro}

The galaxies that populate the present-day Universe, such as our own Milky Way (MW), are predicted to result from the hierarchical assembling of lower-mass progenitor galaxies, some of which formed $>13$ billion years ago \citep[e.g.][]{White1978, Tumlinson2010, Salvadori2010}. During the last decade, our knowledge of the MW in its present state has tremendously advanced thanks to local stellar surveys (e.g., Gaia, \citealt{vallenari2023}, GALAH, \citealt{Buder2021}). Furthermore, observations of ancient metal-poor stars \citep[e.g.][]{Bonifacio2021} are providing key insights on the early stages of the MW evolution, which can be used to indirectly study the formation of the MW and constrain cosmological models \citep[e.g.][]{Hartwig2022, Koutsouridou2023}. However, we lack an understanding of how to link our knowledge of present-day galaxies with those observed at high-z.

The exceptional capability of the \textit{James Webb Space Telescope} (JWST) has opened the possibility of looking directly at the early stages of galaxy formation. Several surveys (e.g. GLASS, \citealt{treu+22}; JADES, \citealt{eisenstein+23}; CEERS, \citealt{finkelstein+23}) have been designed to observe galaxies at the so-called cosmic dawn (redshift $z \approx 15 - 6$) and many bright galaxies have been already discovered and spectroscopically confirmed at $z > 10$ \citep[e.g.][]{Harikane2023, Harikane24, Carniani2024}. Moreover, gravitational lensing enabled to catch the light of fainter and lower-mass galaxies that would otherwise be inaccessible in the early Universe \citep[e.g.][]{roberts-borsani+23, vanzella+23}. 

Ultimately, we are accumulating large amounts of data for both present-day and high-$z$ galaxies. Still, we need to connect these two regimes to understand the overall galaxy-formation process. 

Very recently, the JWST CAnadian NIRISS Unbiased Cluster Survey program (CANUCS, \citealt{willott+22}) has provided a unique opportunity to fill this gap by reporting the discovery of a strongly lensed system at $z=8.3$, the \textit{Firefly Sparkle} \citep{mowla+24}. The Firefly Sparkle comprises 10 systems identified as ``star clusters" and it has two neighboring galaxies at $\approx 2$~kpc and $\approx 13$~kpc. By comparing the stellar masses derived for these systems with those predicted by abundance matching methods, the authors suggest to have observed a MW-like assembly.

However physical properties of distant galaxies, such as the Firefly Sparkle systems, are inferred from photometric data through spectral energy distribution (SED) fitting. 
Inferring the star-formation histories (SFH) of galaxies becomes particularly challenging at high-$z$ due to their bursty nature.
Indeed, both simulations \citep[e.g.][]{Sun23a, Pallottini23, Gelli23} and recent JWST observations \citep[e.g.][]{Endsley23, Langeroodi24, Looser23b} are revealing that in the first billion years of the Universe, galaxies undergo complex and highly time-variable evolution, difficult to capture through SED fitting techniques.
Observing the building blocks of the MW at high-$z$ might allow us to solve this issue. In this case, indeed, we can interpret the observed SEDs by exploiting locally-calibrated MW-formation models to obtain physically-founded SFHs.

In this letter, we interpret the Firefly Sparkle observation using a forward modeling approach based on the predictions of NEFERTITI, a data-calibrated model for the MW formation \citep{Koutsouridou2023}. We aim to answer the following questions: Is the Firefly Sparkle consistent with a MW-like analog at $z\approx 8$? And if so, what are the properties of the observed MW progenitors? By comparing our predicted SEDs with those observed in the Firefly Sparkle we will link the low- and high-$z$ Universe, also providing key predictions to unveil other infant MW-like galaxies with the JWST.

\section{Methods}
\label{sec:methods}

\begin{figure*}[t]
    \includegraphics[width=0.515\hsize]{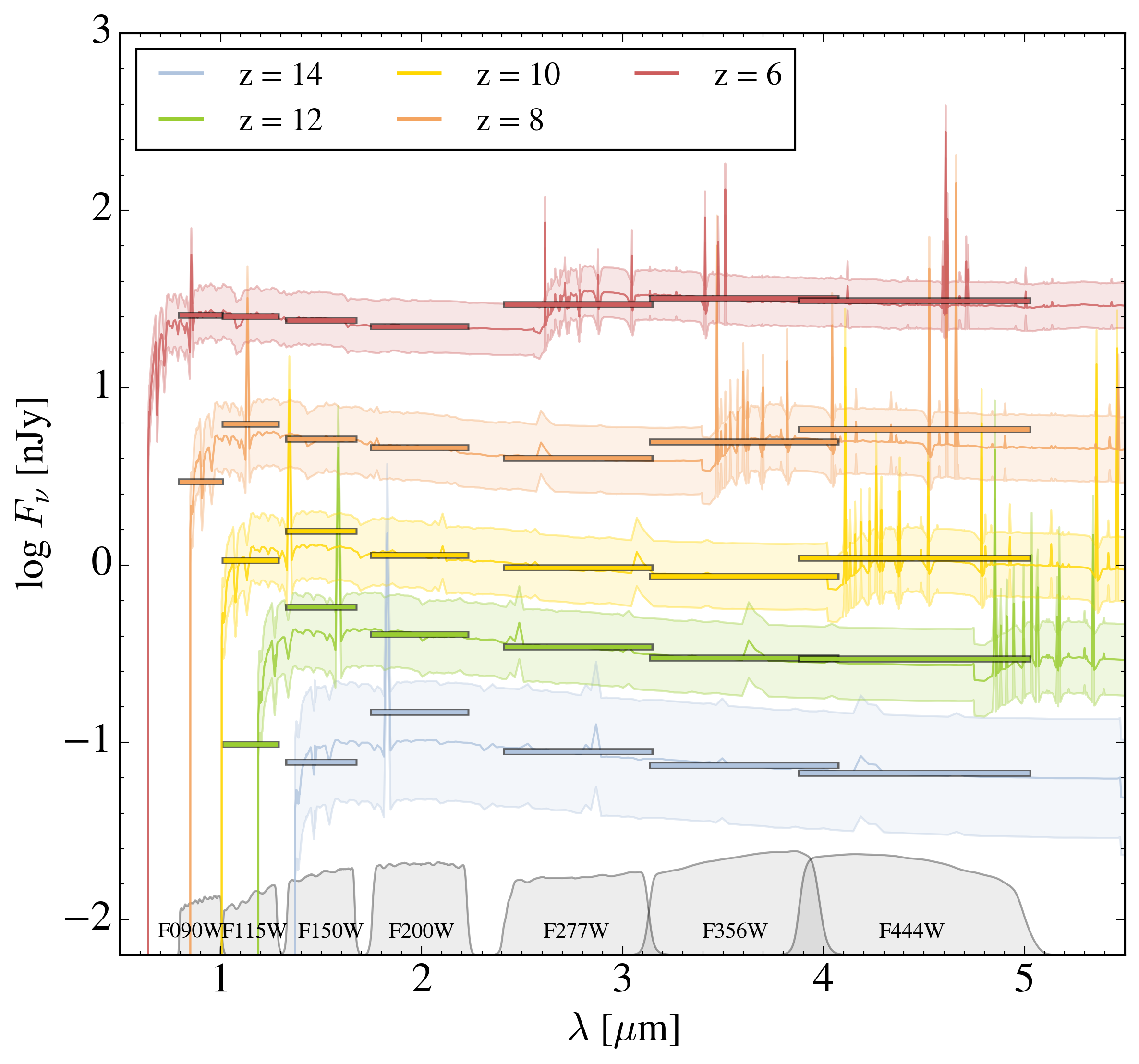} 
    \hspace{0.01\hsize}
    \includegraphics[width=0.485\hsize]{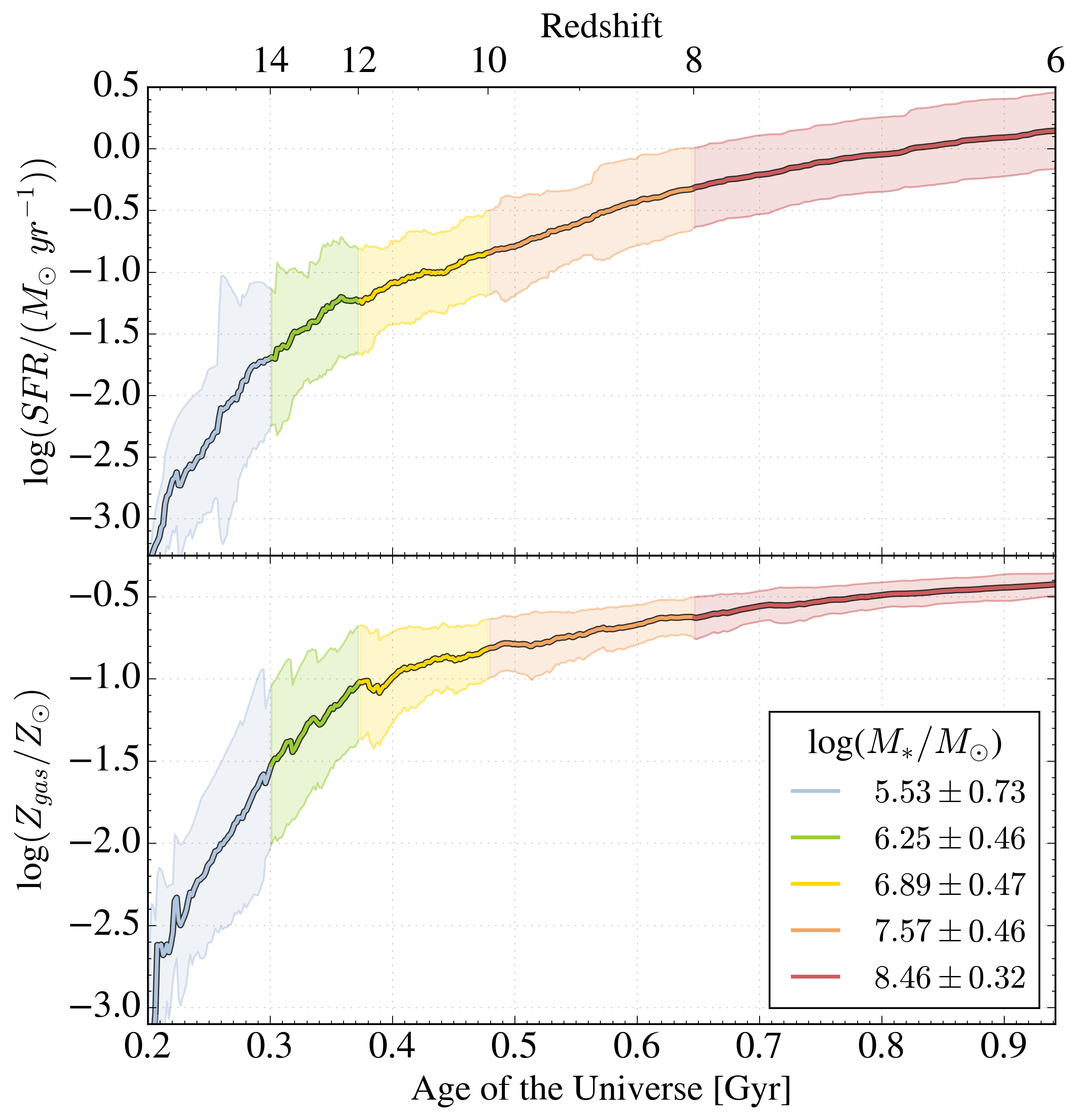} 
    \caption{{\it Left:} Average synthetic SEDs for the MW major branch simulated by the NEFERTITI model, color-coded for redshift. For each mean SED, we show the corresponding photometries in the JWST NIRCam filters. The shaded area represents the standard deviation due to the different merger tree realizations.  {\it Right:} Average star-formation (top panel) and metal-enrichment histories (bottom panel) of the MW major branch, with the same color coding. On the bottom-right, we report the mean $ M_\star$ at the redshifts shown on the left.}
  \label{fig:MWmb}
\end{figure*}

The forward modeling approach consists of two steps: i) predicting the star-formation and chemical evolutionary histories of the high-$z$ building blocks of the MW, i.e. the {\it MW progenitors}, using a data-constrained model; and ii) constructing the synthetic SEDs. In \ref{sec:2.1} we summarize the model employed and in \ref{sec:2.2} we explain how we infer the mock SEDs. 

\subsection{The Milky Way evolution model: NEFERTITI}
\label{sec:2.1}

We employ NEFERTITI (NEar FiEld cosmology: Re-Tracing Invisible TImes, \citealt{Koutsouridou2023}), a state-of-the-art semi-analytical model for the formation and evolution of a MW-like galaxy that can be coupled with N-body simulations of dark matter (DM) only or with merger trees produced via Monte Carlo algorithms. 

A MW analog is defined by the following physical properties of our Galaxy at $z=0$: the virial mass $M_{vir} = (1.3 \pm 0.3) \times 10^{12} \, M_\odot$, the stellar mass $M_\star = (5 \pm 1) \times 10^{10} \, M_\odot$, and the star formation rate $SFR = 1-3 \, M_\odot yr^{-1}$ \citep[e.g.][]{Bland2016}, the metallicity distribution function of halo stars \citep{Bonifacio2021}, the gas-to-stellar mass ratio $M_g/M_\star = 0.1-0.15$ \citep[e.g.][]{Ferriere2001}, and the metallicity of the dense, $Z_{ISM} \approx Z_\odot$, and diffuse gas, $Z_{IGM} \approx 0.1-0.3 \, Z_\odot$ \citep[e.g.][ for more details and references see Sec 2.3 of \citealt{Koutsouridou2023}.]{Tripp2003}. Indeed, the model is calibrated to reproduce all these present-day observational properties of the MW.

NEFERTITI was developed from earlier semi-analytical models for the Local Group formation \cite[][]{Salvadori2007, Salvadori2015, pagnini+23}. Its main innovations are the treatment of the incomplete sampling of the stellar Initial Mass Function (IMF, see \citealt{rossi2021ultra}), the exploration of the unknown IMF of the first (Pop~III) stars and the energy distribution function of the first supernovae \citep[SNe, see][]{Koustouridou2024}. 

NEFERTITI tracks the evolution of baryonic matter inside the DM halos by accounting for star formation and feedback processes. 
Stars form with a rate proportional to the available gas mass in DM halos that reach a minimum mass value, which changes 
over time according to the evolution of the ionizing and photo-dissociating radiation \citep{salvadori2009}. Pop~III stars initially form in the pristine gas, which is then enriched with metals from stellar winds and SNe. SNe explosions can eject a fraction of gas/metals mass outside of the MW progenitors, thus decreasing 
their star formation and enriching the intergalactic medium (IGM). 
Metals are assumed to be instantaneously mixed and when the metallicity of the interstellar medium reaches a value $Z_{gas} > 10^{-4.5} \, Z_{\odot}$ \citep{deBen2017}, normal (Pop~II) stars form according to a Larson IMF \citep{larson1998early} with $m_\star = [0.1,100]\,  M_{\odot}$ and a peak at $m_{ch} =0.35\, M_{\odot}$.  

Here we use NEFERTITI coupled with 50 different merger histories, to have a valid statistical sample. For each merger tree, there is a most massive progenitor (i.e. the MW major branch) and other $\rm 10^{2-4}$ smaller systems. For example, a typical MW major branch at $z=11$ has stellar mass $M_\star \approx 10^{7} \, M_{\odot}$ and gas metallicity $Z_{gas} \approx 0.1 \, Z_{\odot}$, while the other progenitors can go down to $M_\star \approx 10^2 \, M_\odot$ and $Z_{gas} \approx 10^{-4} \, Z_\odot$. For each of these halos, we have access to the star-formation and metal-enrichment histories sampled every Myr from $z=18$ to $z=6$. 
Here we adopt a Pop~III Larson IMF with $m_\star = [0.8,1000] \, M_{\odot}$ and $m_{ch}=10 \, M_{\odot}$, consistently with stellar archaeology observations \cite[see][]{Koutsouridou2023,Koustouridou2024}. However, our findings are independent of the assumed Pop~III IMF. 

\subsection{Building Up Galaxy Spectra}
\label{sec:2.2}

To make accurate predictions for the emission of the MW progenitors at high-$z$, we build their synthetic SEDs using the spectral synthesis code Yggdrasil \citep{zackrisson+11} in the following manner. For each timestep of $1$~Myr, consistently with the NEFERTITI code, we assume an instantaneous instance of star formation. Both the $M_\star$ formed and the $Z_{gas}$ from which the stars form, are taken from the output of the galaxy-formation model. With these properties as input, the Yggdrasil code provides the rest-frame spectrum of a simple stellar population (SSP) at different ages (see \citealt{zackrisson+11} for more details). 
For Pop II stars, the available IMF is the \citealt{kroupa+01} in the interval $m_\star = [0.1-100]\, M_\odot$ and the SSP are from Starburst99 \citep{leitherer+99, vazquez+05}. For Pop III stars we choose the lognormal IMF option in the range $m_\star = [1-500] \, M_\odot$ with the SSP from \citealt{Raiter+10}. 
The SEDs shown in this work will have a maximal nebular contribution, thus no escape of Lyman continuum photons, and no dust corrections.

To model the synthetic SED of the progenitor galaxy at a chosen redshift, we assemble the emitted rest-frame spectra of all the star-formation bursts that make up its evolutionary history and shift into the observed frame of reference. Fig. \ref{fig:MWmb} (left) shows the average SED evolution of the MW major branch and the corresponding synthetic photometries in the JWST NIRCam filters. On the right, we show the average $SFR$ and $ Z_{gas}$ as a function of the cosmic time. The colors mark the redshifts at which the SED is shown.

\section{Results}
\label{sec:results}

\subsection{Detectability of the MW Major Branch}
\label{sec:3.1}

\begin{figure}[t]
    \includegraphics[width=1.\hsize]{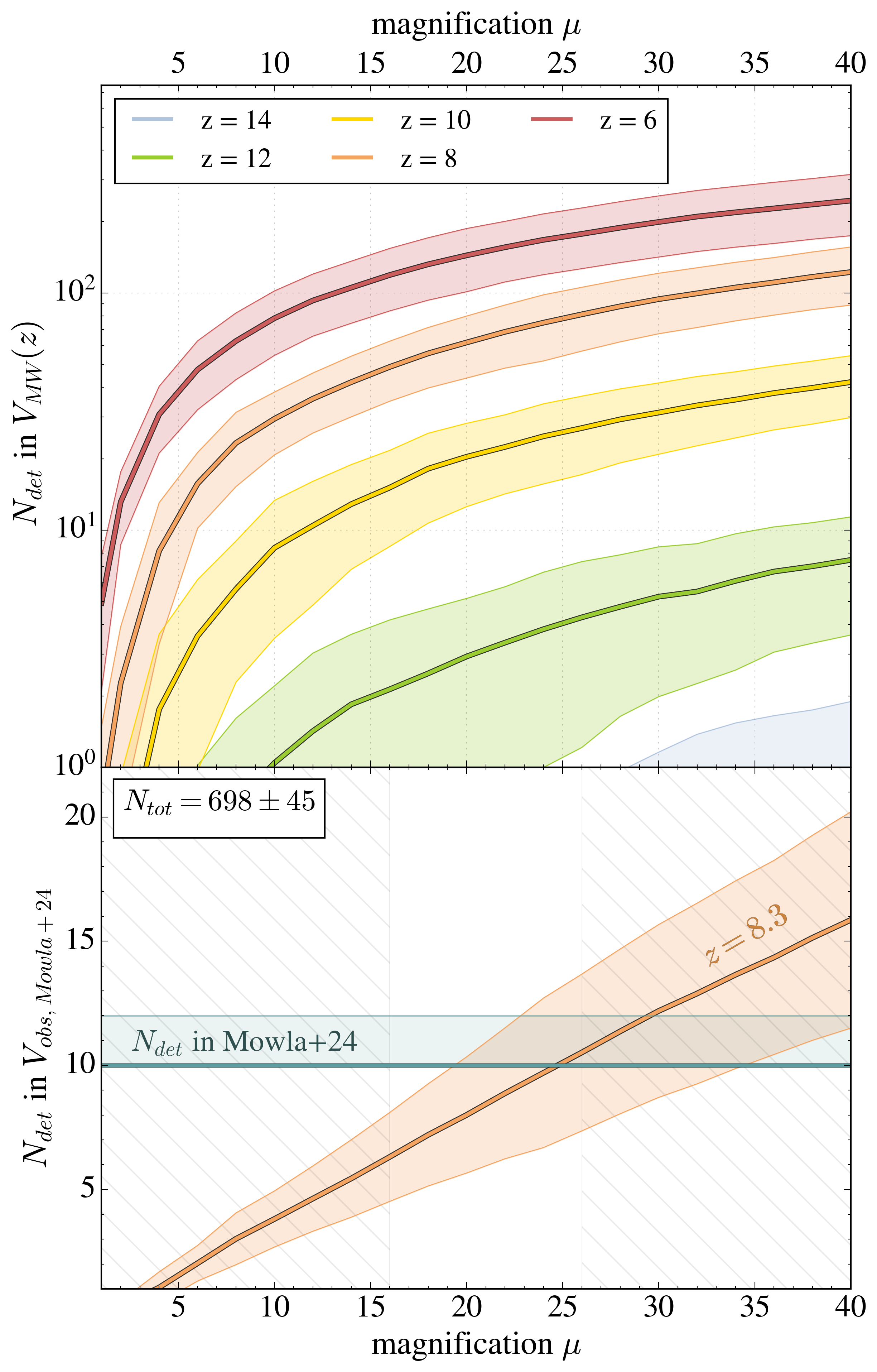} 
    \caption{\textit{Top}: Predictions for the number of detectable progenitors in the virial volume of the MW at different $z$, as a function of the magnification factor. From the 50 merger histories provided by the NEFERTITI model, we calculate the mean number of detections (solid lines) and the standard deviation (shaded areas). \textit{Bottom}: Comparison between the number of systems detected in the Firefly Sparkle (dark green) and the number of progenitors predicted to be detectable in the same conditions (i.e. $z$, volume, lensing). The white area highlights the lensing range of the central region of the Firefly Sparkle.}
  \label{fig:LGdet}
\end{figure}

We investigate the detectability of the most massive progenitor of the MW at different redshifts with JWST NIRCam photometry. To this end, we select the major branch (MB) in each of the 50 merger trees simulated with NEFERTITI (see Sec.~\ref{sec:2.1}). More details on the possible evolution of a MB are in Appendix \ref{appendix:a}.
By modeling their SED evolution as shown in the example in Fig. \ref{fig:MWmb}, we can analyze the possible behavior of a typical MW-like galaxy at high-$z$. We determine the detectability by comparing the synthetic photometry of the MW major branches at different ages (using steps of $\Delta z = 0.1$) to the detection limits of typical deep JWST surveys in the NIRCam wide filters.
Since we do not include the IGM attenuation in our spectral synthesis, we ignore the detectability in the filters at $ \lambda < \lambda_{obs} ( {\rm Ly\alpha}) \approx 1215 \, (1+z) \, \rm \AA $.

Considering the detection limits of the JADES survey (assuming the $5 \sigma$ flux depths from \citealt{rieke+23}), we predict that out of the 50 MW major branch realizations, one is detectable at $z=11$, half at $z=8.2$ and all at $z=7.1$. From this simple analysis, we conclude that the probability for a MW-like progenitor to become detectable in a deep JWST survey like JADES is $P_{det}^{MB}(z<11) \geq 2\%$, $P_{det}^{MB}(z<8.2) \geq 50\%$, and $P_{det}^{MB}(z<7.1)=100\%$

\subsection{Detectability of a MW-type Assembly}

\begin{figure*}
\center

    \includegraphics[width=1.\hsize]{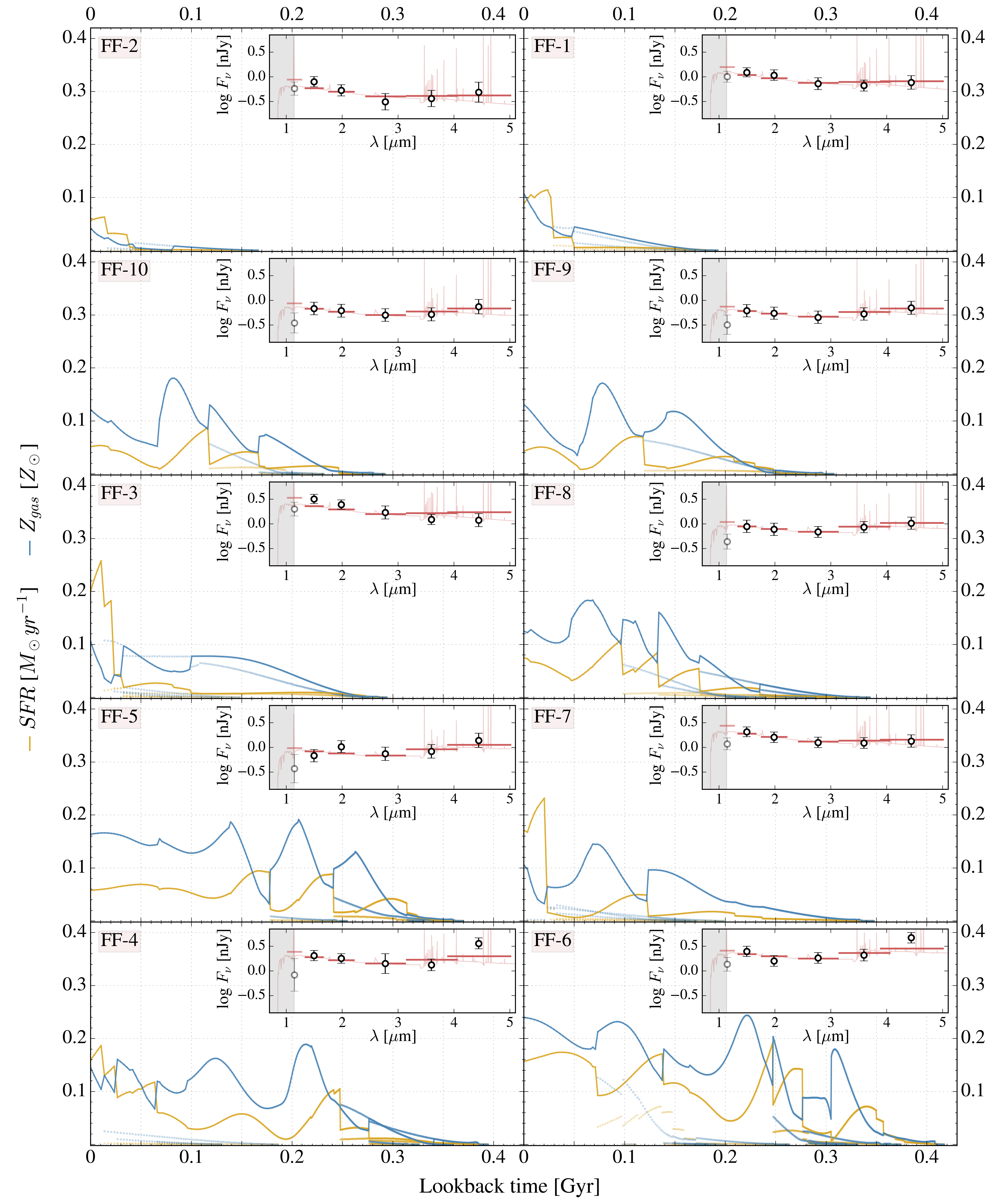}
    \caption{Star-formation (gold) and metal-enrichment (blue) histories of the progenitors that best fit the emission of each system of the Firefly Sparkle, in order of increasing estimated $M_\star$. The synthetic SED is shown in the top-right inset (red), together with the \citealt{mowla+24} observations (black). The grey areas are not included in the fit. The histories of the minor branches (dotted lines) that have merged with the most massive progenitor (solid lines) to form the final one at $z=8.3$ are also shown.}
  \label{fig:fits2}
\end{figure*}

We have shown that a typical MW progenitor can become detectable and appear in JWST observations from $z\lesssim 8$. However, to interpret an observed galaxy as a MW-like progenitor we need to make further predictions on its expected environment and properties. 
Indeed, we expect the most massive MW progenitor at high-$z$ to be surrounded by hundreds of smaller galaxies, which will accrete and merge over time.
Due to their lower masses and weaker emission, we likely need to rely on gravitational lensing to be able to observe such a group of smaller MW building blocks. Here, we investigate their detectability in a typical JWST lensing survey.

By applying the method used in Sec.~\ref{sec:3.1} for the most massive progenitor to {\it all} the MW progenitors predicted by NEFERTITI, we predict the number of MW building blocks that we expect to be detectable for different observational conditions, i.e. as a function of redshift ($z$), observed volume ($V_{obs}$), and magnification factor ($\mu$).
The threshold for detectability is set according to the JWST NIRCam detection limits of the CANUCS program \citep{willott+22}.

Fig. \ref{fig:LGdet} (top panel) shows our predictions at different $z$ for the number of MW progenitors detectable in a JWST lensing survey as a function of $\mu$, within the MW volume, which can be estimated as:

\begin{equation}
    V_{MW}(z) = \frac{4}{3} \pi \left( \frac {R_{MW}^{vir}(0)}{(1+z)} \right )^3
\end{equation}

where $R_{MW}^{vir}(0) \approx  240 \,\rm kpc$.
Without gravitational lensing (i.e. $\mu=1$) we can detect only one progenitor at $z = 8$, as estimated in Sec.~\ref{sec:3.1}, and up to $\approx 5$ MW progenitors at $z = 6$. In general, we see an increasing number of detections both for bigger $\mu$ and for lower $z$, understanding the importance of lensing when aiming to observe the MW assembly at high-$z$.

Our predictions can be easily applied and compared to specific JWST observations to understand if they can be interpreted as MW-like systems, by simply rescaling the MW volume to the one observed:
\begin{equation}
    N_{det}(z)= N_{det, MW}(z) \frac{V_{obs}(z)}{V_{MW}(z)}.
\end{equation}

Let us apply this to the Firefly Sparkle observation.
For the observed region $V_{obs}$ we assume a sphere with $R_{obs}=13~\rm kpc$, corresponding to the estimated distance of the second-closest galaxy to the Firefly Sparkle.

\begin{figure}[t]
\center
    \includegraphics[width=1.\hsize]{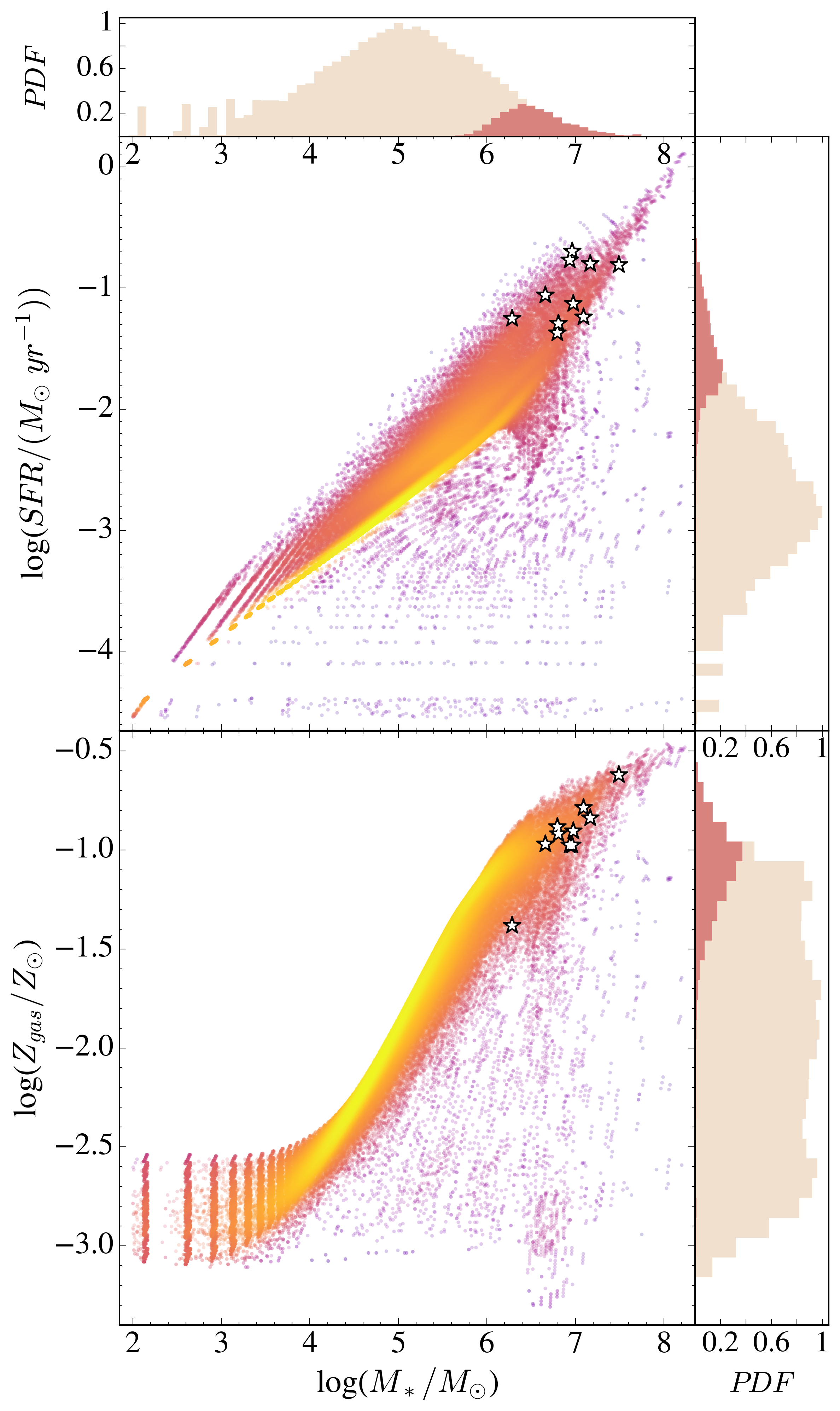}  
    \caption{ $SFR$ (top panel) and $Z_{gas}$ (bottom panel) as a function of $M_{\star}$ for all the MW progenitors predicted by the NEFERTITI model at $z=8.3$, colored by density of points. The best-fitting progenitors to the Firefly Sparkle are highlighted with star markers. On top and the sides, we show the probability distribution functions (PDF) of the properties of all the progenitors (beige) and the detectable ones (red), assuming $\mu=24$ and the detection limits of CANUCS \citep{willott+22}}
  \label{fig:properties}
\end{figure}

In Fig.~\ref{fig:LGdet} (bottom) we show the comparison between the number of systems detected by \citealt{mowla+24} (grey), consisting of 10 systems in the central region and two nearby galaxies, and our predictions for the same observational conditions (orange).
We find that in the magnification range observed in the Firefly Sparkle, $\mu=16-26$, our predicted number of detectable MW progenitors is consistent with those observed.
Although this is just a rough estimate, the remarkable consistency of our predictions with the observed number of sources provides a solid groundwork for their interpretation as MW-like building blocks, i.e. galaxies.
In the next section, we will compare the systems of the Firefly Sparkle one by one to our modeled galaxies.

\subsection{Interpreting the Firefly Sparkle}
\label{sec:3.3}

Here we present the properties of the individual Firefly Sparkle systems inferred through forward modeling with the MW galaxy-formation model NEFERTITI. 
We construct the synthetic SEDs of the MW building blocks according to their star-formation and metal-enrichment histories (see Sec.~\ref{sec:2.2}). 

With this method, we build $\approx 35000$ mock JWST NIRCam photometric data sets, representing 50 different possible populations of MW progenitors at $z=8.3$. Among them, we select the 10 galaxies whose synthetic photometry best fits the observational data of the 10 Firefly Sparkle systems. To this end, we assume the $\mu$ estimated in \citealt{mowla+24} for each ``star cluster" to retrieve its unlensed photometry and compare it to our synthetic SEDs, selecting the one with the minimum reduced $\chi^2$ value. We only consider the central Firefly Sparkle, since we don't have the photometric data of the two nearby galaxies.

Fig.~\ref{fig:fits2} shows the star formation and metal enrichment histories for the ten best-fitting MW progenitors, from the least (top-left) to the most (bottom-right) massive. The insets show the correspondent synthetic SEDs, compared to the photometric JWST data.
The emission of the MW building blocks predicted by NEFERTITI at $z=8.3$ is in good agreement with the photometric data of the Firefly Sparkle, with $\chi_r^2 < 1.4$. We notice that the SFHs of the MW progenitors are not smooth and regular but rather stochastic and characterized by multiple bursts and drops of star formation due to feedback processes (see Sec.~\ref{sec:2.1}). These realistic complex star formation and metal enrichment histories can fit remarkably well the observed emission of all the observed $z=8.3$ systems {\it simultaneously}, proving that we can interpret the Firefly Sparkle as a MW-like assembly.

\begin{deluxetable*}{ccccccc}
\setlength{\tabcolsep}{10pt}
\tablecaption{Properties of the Firefly Sparkle systems inferred via forward modeling with NEFERTITI. 
\label{table:1}}
\tablehead{
\colhead{Name} & \colhead{ $log(M_\star)$} & \colhead{$ log(SFR)$ } & 
 \colhead{ $log(Z_{gas})$  } & \colhead{ $log(M_h)$}  & \colhead{ $t_{half}$ }  & \colhead{ $ min \, \chi^2_{r}$}  \\
 \colhead{\,} & \colhead{$[M_{\odot}]$} & \colhead{$[M_{\odot}\, yr^{-1}]$} & \colhead{$[Z_{\odot}]$} & \colhead{$[M_{\odot}]$} & \colhead{$[Myr]$} & \colhead{\,}  }
\startdata
FF-1 & $6.64 \pm 0.09$ & $-1.06 \pm 0.13$ & $-0.97 \pm 0.15$ & $8.35 \pm 0.08$ & $169 \pm 45$ & 0.232 \\
FF-2 & $6.26 \pm 0.10$ & $-1.25 \pm 0.16$ & $-1.38 \pm 0.14$ & $8.14 \pm 0.08$ & $150 \pm 47$  & 0.340 \\
FF-3 & $6.95 \pm 0.06$ & $-0.70 \pm 0.10$ & $-0.98 \pm 0.11$ & $8.60 \pm 0.04$ & $267 \pm 37$  & 1.367 \\
FF-4 & $7.16 \pm 0.07$ & $-0.80 \pm 0.06$ & $-0.84 \pm 0.06$ & $8.74 \pm 0.05$ & $309 \pm 31$  & 0.900 \\
FF-5 & $7.08 \pm 0.08$ & $-1.24 \pm 0.14$ & $-0.79 \pm 0.14$ & $8.68 \pm 0.07$ & $220 \pm 31$  & 0.407 \\
FF-6 & $7.48 \pm 0.03$ & $-0.81 \pm 0.04$ & $-0.62 \pm 0.03$ & $8.98 \pm 0.02$ & $288 \pm 31$  & 0.809 \\
FF-7 & $6.93 \pm 0.09$ & $-0.77 \pm 0.09$ & $-0.98 \pm 0.07$ & $8.57 \pm 0.07$ & $320 \pm 43$  & 0.092 \\
FF-8 & $6.96 \pm 0.07$ & $-1.13 \pm 0.16$ & $-0.90 \pm 0.21$ & $8.64 \pm 0.06$ & $241 \pm 35$  & 0.002 \\
FF-9 & $6.78 \pm 0.07$ & $-1.37 \pm 0.27$ & $-0.88 \pm 0.27$ & $8.48 \pm 0.05$ & $199 \pm 37$  & 0.027 \\
FF-10 & $6.80 \pm 0.07$ & $-1.29 \pm 0.24$ & $-0.92 \pm 0.24$ & $8.50 \pm 0.05$ & $180 \pm 35$  & 0.057 \\
\enddata 
\end{deluxetable*}
Table \ref{table:1} reports the physical properties inferred for the 10 Firefly Sparkle systems, which in our interpretation are MW-progenitor galaxies. Their stellar masses are $M_\star=10^{6.2-7.5} \, M_{\odot}$, their DM halos have $M_h = 10^{8.1-9} \, M_{\odot}$, the instantaneous $SFR = 0.04 - 0.20 \, M_{\odot} yr^{-1}$, and $Z_{gas} = 0.04 - 0.24 \, Z_{\odot}$. We also report $ t_{half}$, defined as the time needed to form half of the total stellar mass, which is of the order of $150-300$~Myr. As a result, these galaxies have old stellar populations that have formed more than 100 Myr before the observation. 

In Fig. \ref{fig:properties} we show the properties of all the $z=8.3$ MW building blocks in 50 realizations of NEFERTITI. 
On the top and right panels, we can visualize the probability distribution of their properties, where we have highlighted in red those of the galaxies that can be detected in the observational conditions of \citealt{mowla+24} and assuming $\mu=24$ (for different $\mu$ see Appendix \ref{appendix:b}). The detectable MW progenitors are the most massive, star-forming, and metal-rich systems, meaning that we are observing only a biased sample. The star markers show the galaxies that best fit the individual Firefly Sparkle systems, which are scattered across the observable regions.

Here we show all the 50 merger tree realizations to visualize the possible properties of a MW-assembly environment at $z=8.3$. If we focus on a single assembling history, only a couple of galaxies are more massive and star-forming than our best fits. They could be the two nearby galaxies, but we need further photometric observations to conclude their interpretation. 

\section{Discussion and Conclusions}
\label{sec: conclusions}

The recent JWST CANUCS observation of the Firefly Sparkle claimed to be a MW-like galaxy assembly at $z\approx 8$, presented us with the unique opportunity to connect the still debated early stages of galaxy formation to our knowledge of the Local Universe. 
In this Letter, we performed forward modeling of the MW progenitors with the galaxy-formation model NEFERTITI to make both predictions and interpretations for JWST. We chose to use Monte Carlo merger trees to have a valid statistical sample of different possible merger histories for the Milky Way.

First, analyzing the multiple realizations of possible MW assembly histories, we find that the most massive MW progenitor can appear within JADES detection limits as early as $z\approx 11$ ($2\%$ probability), it is very likely detectable at $z\approx 8.2$ ($50\%$), and always detectable at $z\leq 7.1$. 
As a result of this prediction, in both completed and ongoing deep surveys of JWST, we could have observed the major branch of MW-like galaxies that still require interpretation. 

Second, we confirm the interpretation of the Firefly Sparkle as a MW analog. The number of systems detected in \citealt{mowla+24} is indeed consistent with the predicted number of detectable MW progenitors assuming the same observational conditions. Moreover, the NIRCam photometric data of the 10 Firefly Sparkle ``star clusters" are in good agreement with the synthetic photometries of the MW progenitors at the same redshift, with $\chi^2_r < 1.4$. By identifying the MW building blocks in NEFERTITI that best fit the observations, we estimate the physical properties of each of these systems, which are not predicted to be star clusters but rather MW progenitor {\it galaxies} with $M_h = 10^{8.1-9} \, M_{\odot}$, $ M_\star = 10^{6.2}- 10^{7.5} \, M_{\odot}$, $ SFR = 0.04-0.20 \, M_{\odot} yr^{-1}$ and $ Z_{gas} = 0.04 - 0.24 \, Z_{\odot}$.

The stellar masses inferred by \citealt{mowla+24} with spectrophotometric fitting techniques \citep{iyer+17, iyer+19}, $M_{\star, obs} = 10^{5.3-5.8} \, M_\odot$, are on average lower than our predictions. This is not surprising given the bursty and long SFHs of our MW progenitors, with $\rm t_{half}=150-300$~Myr. Indeed, reconstructing and resolving old stellar populations through SED fitting is particularly difficult due to the outshining of their stellar emission by the dominating nebular emission from young stars \citep[e.g.][]{Whitler23, Gimenez-Arteaga24}.
However, if we self-consistently calculate the $SFR$ values from the specific $SFRs$ reported in the Firefly Sparkle paper, and their respective errors, we find $SFR_{obs} = 0.003 - 0.6 \, M_\odot yr^{-1}$, which are consistent with our estimates.

As for the gas metallicity, \cite{mowla+24} derive a single value for the entire system, $Z_{gas, obs}=0.02 \pm 0.01 \, Z_{\odot}$. We find the $ Z_{gas}$ and its evolution for each MW progenitor, with values that are higher than their estimation but still subsolar.
The reason for this discrepancy could be due to the direct $ T_e$ method used to infer $Z_{gas}$ from the observed spectrum. By fitting the observations with the HOMERUN code \citep{marconi+24}, which uses a weighted combination of multiple single-cloud photoionization models, we found a total metallicity value that is consistent with those predicted by the NEFERTITI model. 
 
In conclusion, our results show that the Firefly Sparkle observation is consistent with what we predict for a MW-like environment at $z=8.3$. Our forward modeling approach 
has allowed us to estimate the properties of the individual MW building blocks with physically motivated and complex evolutionary histories. To help ongoing and future JWST surveys understand if they are witnessing high-$z$ MW-like environments in lensed fields, we provide predictions for the number of detectable MW progenitors at various redshifts and for different magnification factors. Catching other MW-like systems at cosmic dawn will indeed allow the community to collect different pictures of a MW-analog during its infancy. Ultimately, more JWST observations combined with our predictions might allow us to link the high- and the low-$z$ perspectives to understand the early assembly stages of our Galaxy.

\section*{Acknowledgements}

This project received funding from the ERC Starting Grant
NEFERTITI H2020/804240 (PI: Salvadori).\\

\bibliography{JWST_progenitors,codes}
\bibliographystyle{aasjournal}

\appendix
\section{Properties of the MW major branches}
\label{appendix:a}
\begin{figure} [h]
\centering
    \includegraphics[width=1.\hsize]{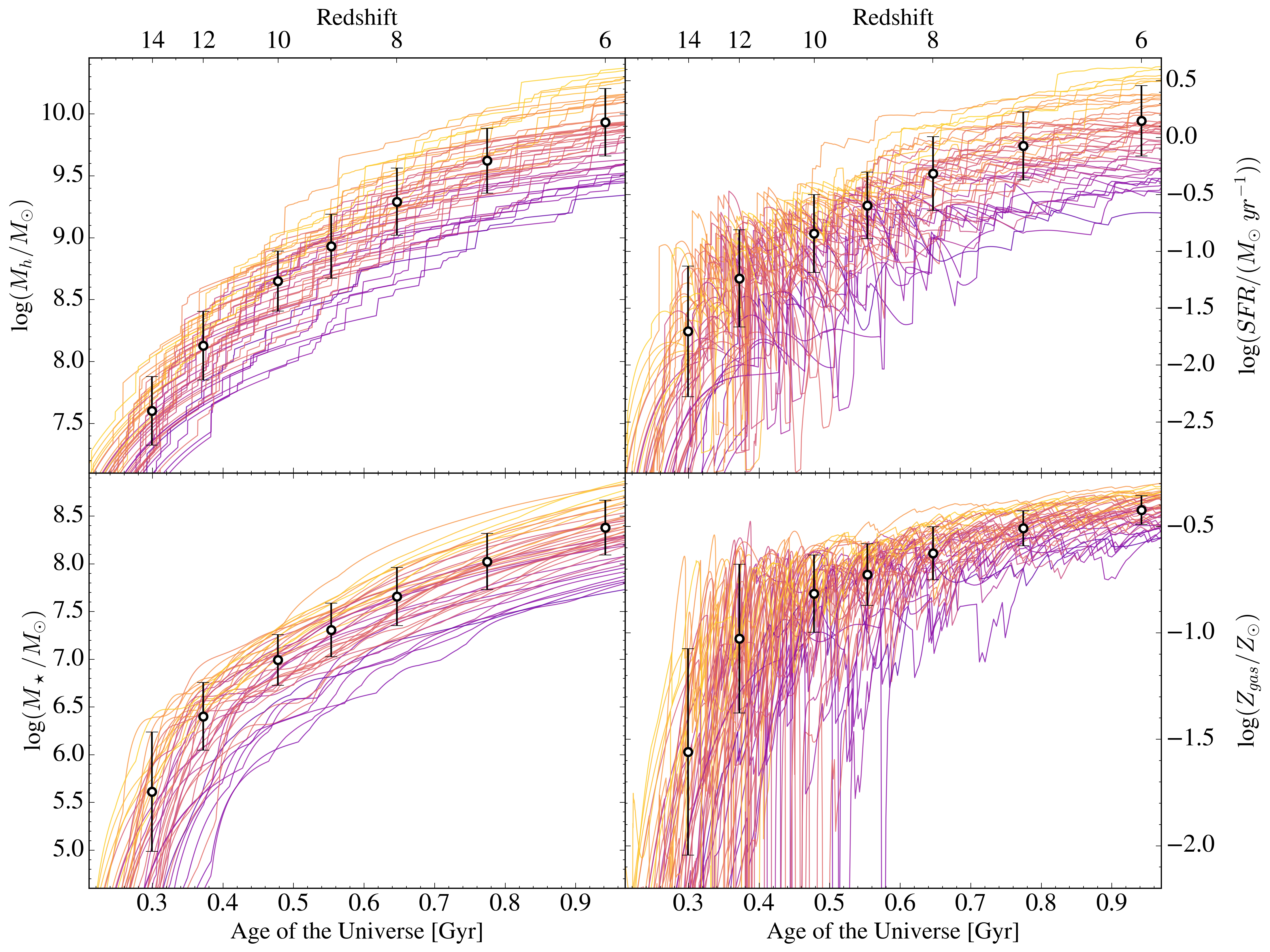} 
    \caption{ Physical properties of all the MW major branches simulated by NEFERTITI: halo mass (top left), stellar mass (bottom left), star formation rate (top right), and gas metallicity (bottom right), colored by halo mass at $z=6$. The black error bars represent the mean value and the relative standard deviation at $z=14, 12, 10, 9, 8, 7, 6$.}
  \label{fig:MWm}
\end{figure}

The 50 merger tree realizations of the NEFERTITI model provide us with 50 different possible formation histories of the MW major branches. Fig. \ref{fig:MWm} shows the evolution of the physical properties (i.e. $M_h$, $M_\star$, $SFR$, $Z_{gas}$) for each merger tree's most massive MW progenitor, from $z \approx 18$ to $z \approx 6$.
The wide range of possible evolutions showcases the importance of using the DM merger tree method \citep[e.g. see][for more details]{Salvadori2007}, which allows us to have a significant statistical sample. We can notice how the $SFR$ evolves in a stochastic and bursty manner due to the stellar feedback \citep[see Sec.~\ref{sec:2.1} and][for more details on NEFERTITI]{Koutsouridou2023} and to the galaxy merger process, clearing once more the importance of using physically-motivated SFHs when interpreting JWST observations at high-$z$.

\section{Detectable MW progenitors with different lensings}
\label{appendix:b}

\begin{figure}[h]
\center

    \includegraphics[width=0.5\hsize]{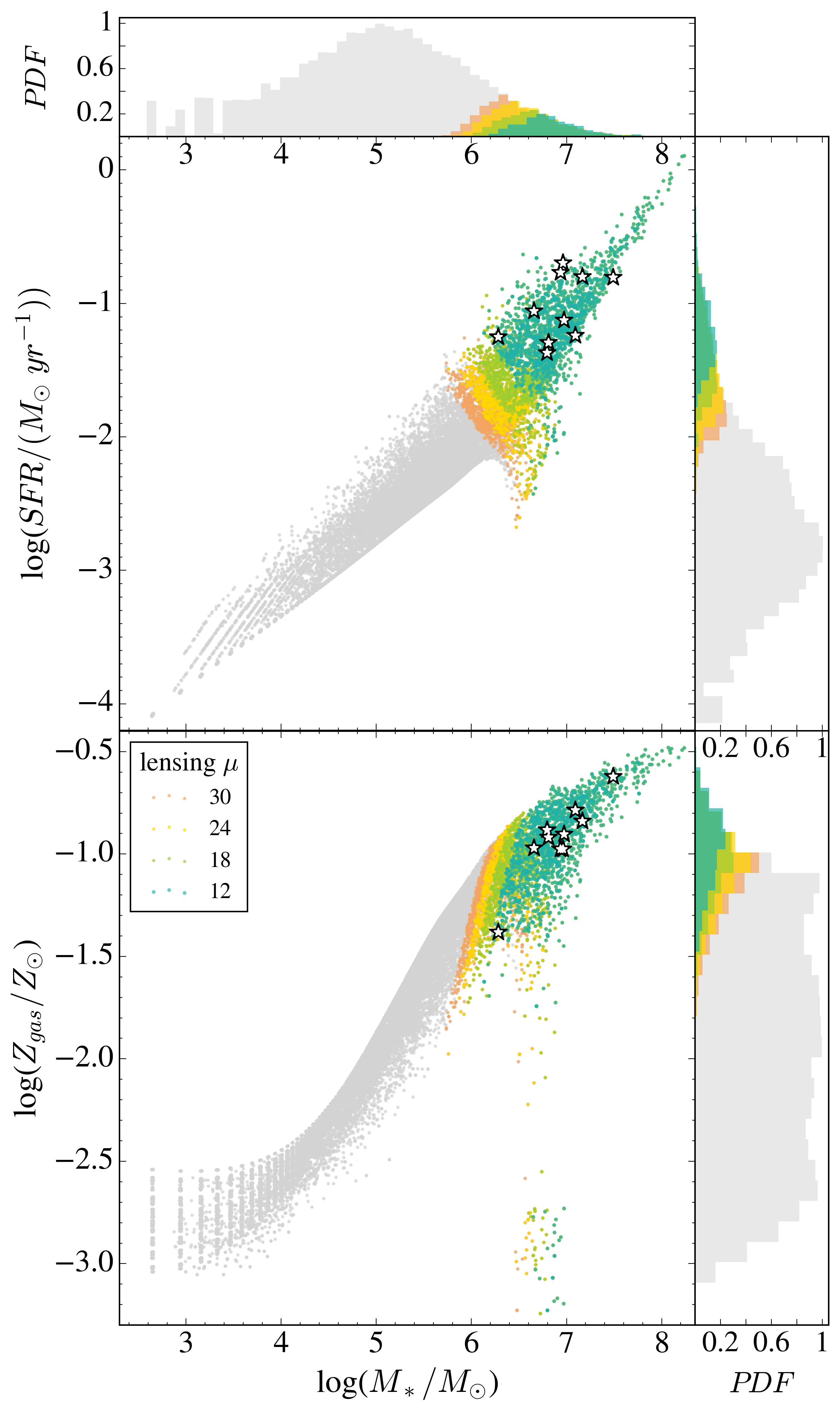}  
    \caption{ $SFR$ (top panel) and $ Z_{gas}$ (bottom panel) as a function of $\ M_{\star}$ for all the MW progenitors predicted by the NEFERTITI model at $z=8.3$. Assuming the detection limits of CANUCS \citep{willott+22}, the grey points represent the non-detectable systems, while the colored ones are detectable with different magnification factors (i.e. orange for $\mu = 30$, yellow for $\mu = 24$, light green for $\mu=18$ and dark green for $\mu=12$). The best-fitting progenitors to the Firefly Sparkle are highlighted with star markers. On top and the sides, we show the probability distribution functions (PDF) of the properties of all the progenitors and of the detectable ones, again color-coded for $\mu$. }
  \label{fig:properties2}
\end{figure}

In Sec. \ref{sec:3.3} we found that the MW-progenitor galaxies that can be detected in a JWST lensing survey like CANUCS \citep{willott+22} are in the high-tail of the $M_\star$, $SFR$, and $Z_{gas}$ distributions. To better visualize the detectable galaxies, in Fig. \ref{fig:properties2} we show the final properties of all the MW progenitors at $z=8.3$ for 50 merger histories, with each point representing a progenitor galaxy. In this way, we can highlight the single galaxies that can be detected at different $\mu$. We notice that there are $SFR-M_\star$ and $Z_{gas}-M_\star$ relations for the detectability: 
\begin{equation}
    log(SFR) \gtrsim - log(M_\star) + a(\mu)
\end{equation}
\begin{equation}
    log(Z_{gas}) \lesssim 2 \, log(M_\star) + b(\mu).
\end{equation}

These relations quantify how we can observe lower-mass systems if they are more star-forming or metal-poor, and how a higher magnification factor can help us access more typical MW progenitor galaxies.

\end{document}